\documentclass[12pt,leqno]{article}
\usepackage{latexsym}
\newtheorem{theorem}{Theorem}
\newtheorem{corollary}{Corollary}
\newtheorem{lemma}{Lemma}

\newtheorem{definition}{Definition}

\newcommand{\blackslug}{\mbox{\hskip 1pt \vrule width 4pt height 8pt 
depth 1.5pt \hskip 1pt}}
\newcommand{\qed}{\quad\blackslug\lower 8.5pt\null\par\noindent}
\newenvironment{proof}{\par\noindent{\bf Proof:}}{\qed \par}

\newcommand{\cC}{\mbox{${\cal C}$}}
\newcommand{\cH}{\mbox{${\cal H}$}}

\newcommand{\cR}{\mbox{${\cal R}$}}

\newcommand{\eqdef}{\stackrel{\rm def}{=}}

\title{Quantic Superpositions and the Geometry of Complex Hilbert Spaces
\thanks{This work was partially supported 
by the Jean and Helene Alfassa fund for 
research in Artificial Intelligence, by the Israel Science Foundation grant 
183/03 on ``Quantum and other cumulative logics'' and by EPSRC Visiting 
Fellowship GR/T 24562 on ``Quantum Logic''}
}

\author{Daniel Lehmann\\Selim and Rachel Benin School of Computer Science and
Engineering\\and Center for the Study of Rationality
\\Hebrew University, \\Jerusalem 91904, Israel
}
\date{July 2007}

\begin{document}
\maketitle
\begin{abstract}
The concept of a superposition is a revolutionary novelty introduced
by Quantum Mechanics.
If a system may be in any one of two pure states $x$ and $y$, we must consider
that it may also be in any one of many {\em superpositions} of $x$ and $y$.
An in-depth analysis of superpositions is proposed, in which states are 
represented by one-dimensional subspaces, not by unit vectors as in Dirac's
notation.
Superpositions must be considered when one cannot distinguish
between possible paths, i.e., histories, leading to the current state of the
system. In such a case the resulting state is some compound of the states
that result from each of the possible paths. States 
can be compounded, i.e., superposed in such a way only 
if they are not orthogonal. Since different classical states are orthogonal,
the claim implies no non-trivial superpositions can be observed in classical
systems. 
The parameter that defines such compounds is a
proportion defining the mix of the different states entering the compound.
Two quantities, $p$ and $\theta$, both geometrical in nature, relate
one-dimensional subspaces in complex Hilbert spaces: the first one is a measure
of proximity relating two rays, the second one is an angle relating three rays.
The properties of superpositions with respect to those two quantities 
are studied.
The algebraic properties of the operation of superposition are very different
from those that govern linear combination of vectors.
\end{abstract}

\section{Introduction and Previous Work} \label{sec:intro}
During the elaboration of~\cite{BirkvonNeu:36} John von Neumann wrote
to Garret Birkhoff: ``I would like to make a confession
which may seem immoral: I do not believe absolutely in Hilbert space any more. 
After all Hilbert-space (as far as quantum-mechanical things are concerned) 
was obtained by generalizing Euclidean space, footing on the principle of 
``conserving the validity of all formal rules''. 
This is very clear, if you consider the axiomatic-geometric definition of
Hilbert-space, where one simply takes Weyl's axioms for a unitary-Euclidean
space, drops the condition on the existence of a finite linear basis, and
replaces it by a minimum of topological assumptions 
(completeness + separability). Thus Hilbert-space is the straightforward
generalization of Euclidean space, if one considers the {\em vectors}
as the essential notions. Now we begin to believe that it is not the 
{\em vectors} which matter but the {\em lattice of all linear (closed) 
subspaces}. Because:
\begin{enumerate}
\item The vectors ought to represent the physical {\em states}, but they
do it redundantly, up to a complex factor only.
\item And besides the {\em states} are merely a derived notion, the primitive
(phenomenologically given) notion being the {\em qualities}, which correspond 
to the {\em linear closed subspaces}'' 
(see~\cite{vNeumann_letters}, p. 59, letter dated Nov. 13, Wednesday, 1935).
\end{enumerate}

The goal of this work is to pursue von Neumann's program of describing 
Quantum Logic in terms of closed subspaces and without vectors one step
further. 
This work presents two original features:
\begin{itemize}
\item it takes a logical approach to Quantum Physics, where states and
propositions take the main roles, and
\item while it assumes the formalism of Hilbert spaces that fits Quantum 
Physics, it tries the utmost to use only notions, such as states, propositions,
projections, orthogonality and so on, that have a meaning, albeit mostly
trivial, in Classical Physics.
Special care will be taken to ensure that the quantic principles proposed
hold classically.
\end{itemize}

\section{Quantum Logic} \label{sec:Logic}
One may say that Logic is the study of the relation between states of the 
world and propositions used to talk about those states. Quantum logic must
therefore be the study of the relation between quantum states and quantum
propositions. The accepted view is that both quantum states and quantum
propositions should be represented by closed subspaces of a Hilbert space.
Quantum states are one-dimensional subspaces.
Quantum logic is therefore the study of the relation between one-dimensional 
subspaces and arbitrary closed subspaces.
One obvious topic for Quantum logic is therefore the study of the properties
of projections in Hilbert spaces: a one-dimensional subspace projects onto
a one-dimensional or zero-dimensional subspace of any closed subspace.
Projections are also central to Quantum Physics since they correspond to the 
change brought about by the measurement of a physical property. 
Previous works~\cite{LEG:Malg} and~\cite{AndThen:Leibniz} provided 
a first study of some of the properties of such projections: 
they dealt only with qualitative properties. 
The present paper inaugurates the quantitative study of the 
projective geometry of complex Hilbert spaces.

The purpose of the exercise is to shed light on the notion of measurement 
in Quantum Physics by developing a geometry of Hilbert spaces whose 
entities are physically meaningful: states of physical systems and 
measurements on physical systems. Our goal can be understood in considering 
the history of geometry. Euclidean plane geometry was the starting point. 
Its elements are points and lines. 
Mathematical developments (due to Descartes in particular) 
enabled a treatment of geometry in the vector space
$\cR^{n}$. A new definition of geometry, abstracting from the vector space
structure and returning to the basic notions of points and lines, enabled
the development of non-Euclidean geometries.
For Hilbert spaces, historically the algebraic presentation came first.
The purpose of this paper is to extract from the algebraic presentation
a leaner presentation similar in spirit to Euclid's geometry.
Our basic entities are one-dimensional subspaces and, more
generally, closed subspaces {\em and not vectors}.

In an obvious way, two elements (vectors) of a Hilbert space 
define a number, their inner product. 
We are looking for numbers that characterize relations between subspaces, not 
vectors. 
This paper proposes to associate a real number
with any pair of one-dimensional subspaces $x, y$: \mbox{$p(x, y)$} 
and, by extension,
to any pair of a one-dimensional subspace and a closed 
subspace $\alpha$: \mbox{$p(x, \alpha)$}. 
This number is always in the interval $[0,1]$ 
and behaves in many ways like {\em the probability that the proposition 
$\alpha$ is
found true when it is tested for in state $x$}, in line with the probabilistic
interpretation of Quantum Physics. 
It satisfies further properties that
are more difficult to interpret and that characterize the linear dependence
structure and the structure of projections.

Another numerical quantity, an angle, $\theta$, is defined by any triple of 
one-dimensional subspaces. It is interpreted as the source of the
interference occurring between alternative paths a system could take.
This paper is devoted to the study of those aspects of the geometry of
Hilbert spaces related to the numbers $p$ and $\theta$. 
The study of those M-algebras 
(see~\cite{LEG:Malg}) that admit quantities satisfying the properties of $p$,
$\theta$ and superpositions is left for further study.

\section{Background and Notations} \label{sec:background}
We assume a Hilbert space \cH\ on the field \cC\ of complex numbers is given.
The complex conjugate of a complex number $c$ is $\overline{c}$.
For any complex number $c$, $\mid \! c \! \mid$ 
represents its modulus, which is 
a nonnegative real number.
For any complex number $c$ different from $0$, $\arg(c)$ represents its
complex argument: \mbox{$c = \, \mid \! c \! \mid e^{i \arg(c)}$}.
Elements of \cH\ will typically be: \mbox{$\vec{u}, \vec{v} \ldots$}.
The zero vector is denoted by $\vec{0}$. The inner product of $\vec{u}$
and $\vec{v}$ is \mbox{$\langle \vec{u} \, , \, \vec{v} \rangle$}.
The inner product is linear in its first argument and conjugate-linear in
its second argument.
Two vectors $\vec{u}$ and $\vec{v}$ are perpendicular, written 
\mbox{$\vec{u} \perp \vec{v}$}, iff 
\mbox{$\langle \vec{u} \, , \, \vec{v} \rangle = 0$}.
The norm of $\vec{u}$ is \mbox{$\parallel \vec{u} \parallel$}.
A unit vector is a vector of norm $1$.
We shall use the notation \mbox{$\langle \vec{u} \, , \, \vec{v} \rangle > 0$}
to denote the fact that the inner product is a strictly positive {\em real} 
number.

The set of all closed subspaces of \cH\ will be denote by $M$.
The elements of $M$ should be thought of representing propositions, 
or, results of physical measurements. Greek letters from the beginning
of the alphabet will be used to denote elements of $M$. The reader may
think of a typical element of $M$, $\alpha$ as meaning {\em the spin
in the $z$-direction is nonnegative}. Note that propositions represent
measurements with a specified result or a set of possible results: 
such as measuring the value $1/2$ for
the spin in the $z$-direction or measuring a nonnegative value for this spin.
To every \mbox{$\alpha \in M$}
one may associate its orthogonal complement, which will be denoted
$\neg \alpha$. The proposition $\neg \alpha$ is interpreted as the measurement
that measures the quantity measured by $\alpha$ but provides a value that is
not in the set specified by $\alpha$. If $\alpha$ claims that the spin in
the $z$-direction is nonnegative, $\neg \alpha$ measures the spin along the
same direction but finds it negative. 
Two specific propositions are worth mentioning: falsehood, $0$
is the null subspace $\{\vec{0}\}$ and truth, $1$ is the whole space \cH.
Any closed subspace $\alpha$ of \cH\ defines the projection of \cH\ onto
$\alpha$. For any \mbox{$\vec{u} \in \cH$} its projection on $\alpha$ will
be denoted $\alpha(\vec{u})$. The relation between physical measurements and
projections will be explained after we discuss states.
 
Among the closed subspaces of \cH\ particular attention will be paid to
one-dimensional subspaces. The set of one-dimensional subspaces of \cH\
is denoted $X$ and the elements of $X$ are typically letters from the end
of the alphabet: $x$, $y$ and so on. As mentioned just above:
\mbox{$X \subseteq M$}. Elements of $X$ will be called {\em states}.
A one-dimensional subspace $x$ represents a possible (pure) state of the 
physical system. Think of the state in which the spin in the $z$-direction is 
$1/2$, for example. We assume that states are propositions.
The fact that \mbox{$X \subseteq M$} reflects
the situation in which every pure state has an associated measurement that
characterizes it: one may measure the spin in the $z$-direction and one
of the possible values is $1/2$. The proposition ``the spin in the 
$z$-direction is nonnegative'' is not a state.

Since a proposition \mbox{$\alpha \in M$} is a closed subspace of \cH,
for any \mbox{$x \in X$}, either \mbox{$x \subseteq \alpha$} or $\alpha$ 
contains no vector of $x$ except the zero vector. Any proposition is the union
of the states it includes and any proposition can be seen as the set of all
the states it includes. We shall indeed prefer the notation 
\mbox{$x \in \alpha$} to \mbox{$x \subseteq \alpha$}.

Note that if 
\mbox{$\vec{v} \in x \in X$} and \mbox{$\vec{u} \perp \vec{v}$} then
\mbox{$\vec{u} \perp \vec{w}$} for every \mbox{$\vec{w} \in x$}.
We denote such a situation by \mbox{$\vec{u} \perp x$}.
If every \mbox{$\vec{u} \in \alpha$} is orthogonal to $x$ we say that
\mbox{$x \perp \alpha$}. If every \mbox{$x \in X$}, 
\mbox{$x \in \alpha$} is orthogonal to $\beta$ we say that
\mbox{$\alpha \perp \beta$}.
The image of any \mbox{$x \in X$} by any (projection) \mbox{$\alpha \in M$} 
is either a one-dimensional subspace \mbox{$y \in X$} or the zero-dimensional
subspace. This second possibility occurs exactly when $x$ is orthogonal
to $\alpha$. We shall denote by $\alpha(x)$ the one-dimensional 
or zero-dimensional subspace that is the projection of $x$ onto $\alpha$.
Note that \mbox{$\alpha(x) = x$} iff \mbox{$x \in \alpha$}.
We write \mbox{$\alpha(x) = 0$} to denote the case $\alpha(x)$ 
is zero-dimensional, i.e., the case \mbox{$x \perp \alpha$}.
The projection of the zero-dimensional subspace on any $\alpha$ is the
zero-dimensional subspace and we shall extend the action of $\alpha$ by
setting \mbox{$\alpha(0) = 0$}.

In Quantum Physics measurements may change
the state of the system. The state obtained when measuring $\alpha$ in state
$x$ is precisely $\alpha(x)$, the projection of $x$ on the subspace $\alpha$.
If $x$ is orthogonal to $\alpha$, then the measurement $\alpha$ is impossible
in state $x$: this happens precisely when the quantity measured by $\alpha$
has, in $x$, a well-defined value that is not in the set specified by $\alpha$.
Equivalently, this happens precisely when $x$ is in the subspace $\neg \alpha$,
or \mbox{$(\neg \alpha)(x) = x$}. 

\section{Classical Physics} \label{sec:classical}
The notions described in Section~\ref{sec:background} have been given
a meaning grounded in the Hilbert space formalism of Quantum Mechanics.
This seems to preclude their application to Classical Mechanics, since,
classically, states are not rays in a Hilbert space.
Nevertheless, the common wisdom is that Quantum Mechanics should apply 
everywhere and that Classical Mechanics should be a limiting case of
Quantum Mechanics. Indeed, both Classical Mechanics and Quantum Mechanics
can be studied in structures that abstract from the concepts
of Section~\ref{sec:background}, preserving the properties of states
and measurements. A full treatment is left for future work, 
but the following remark explains the main feature of classical systems.

Classically, measurements do not change the state of a system, therefore if
a state $x$ is not orthogonal to a proposition $\alpha$,
we have \mbox{$\alpha(x) = x$}, expressing the fact that either $x$ possesses
the property $\alpha$ or it possesses its negation $\neg \alpha$. We have:
\[
{\bf Principle \ of \ Classical \ Physics \ } 
{\rm \ Any \ two \ different \ states \ are \ orthogonal.}
\]

\section{The Reciprocity Principle} \label{sec:reciprocity}
Before proceeding to the analysis of the notion of a superposition which is the
crux of this paper, we need a simple remark. It will be presented as a 
principle, to stress the physical meaning of a fact that is woven so deep in 
the familiar linear structure of Hilbert spaces that we tend not to reflect 
on it anymore.
If the measurement $\neg x$ acting on state $y$ and on state $z$ 
produces the same state, then $x$, $y$ and $z$ must sit in the same
plane, and therefore the measurement $\neg y$ must produce the same state
when acting on $x$ and on $z$.
\pagebreak[0]
\[
{\bf Reciprocity \ Principle \ } {\rm \ Let \ } x, y, z \in X, 
{\rm \ be \ pairwise \ different}.
\]
\[
{\rm Then \ } (\neg x)(y) = (\neg x)(z) \: \Rightarrow \: 
(\neg y)(z) = (\neg y)(x).
\]

The Reciprocity Principle suggests the following definition.
\begin{definition} \label{def:coplanarity}
We shall say that states $x$, $y$ and $z$ are {\em coplanar},
written \mbox{$coplanar(x, y, z)$} iff either two out of the three
are equal, or they are pairwise different and 
\mbox{$(\neg x)(y) = (\neg x)(z)$}. 
\end{definition}
The Reciprocity Principle says that 
{\em coplanarity} is a property of the set \mbox{$\{x, y, z\}$}, 
i.e., for any permutation $x'$, $y'$, $z'$ of $x$, $y$, $z$
\mbox{$coplanar(x', y', z')$} is equivalent to \mbox{$coplanar(x, y, z)$}.

The Reciprocity Principle is experimentally testable: if the {\em no} answer
to a test $x$ gives the same state when performed on $y$ and on $z$, the
{\em no} answer on a test $y$ will give the same answer on $z$ and $x$.

In Hilbert space, indeed, if $y$ and $z$ have the same projection on the 
subspace orthogonal to $x$, call it $x'$, then all four one-dimensional
subspaces: $x$, $x'$, $y$ and $z$ are in the same two-dimensional subspace,
call it $\alpha$,
and therefore the projections of $z$ and $x$ on the subspace orthogonal
to $y$ are both the one-dimensional subspace of $\alpha$ orthogonal to $y$.

In Classical Physics, the Reciprocity Principle holds trivially,
since its assumptions are never satisfied. 
Indeed if \mbox{$x \neq y$}, we have 
\mbox{$(\neg x)(y) = y$}, and similarly
\mbox{$(\neg x)(z) = z$} and therefore the assumption
\mbox{$(\neg x)(y) = (\neg x)(z)$} implies \mbox{$y = z$}, contrary to 
assumption.

\section{Superpositions: Conceptual Analysis} \label{sec:conceptual}
The concept of a superposition is a revolutionary novelty introduced
by Quantum Mechanics.
If a system may be in any one of two pure states $x$ and $y$, we must consider
that it may also be in any one of many {\em superpositions} of $x$ and $y$.
This paper is devoted to an in-depth analysis of superpositions.

The following remark has resulted in a vast literature: 
the revolutionary character of quantic superpositions is the consequence
of the fact no such superpositions have to be considered, or may be 
seen in classical systems. In Schr\"{o}dinger's colorful thought experiment:
the cat is either dead or alive, but nobody has evidence of a superposition
of a dead and a live cat. This seems to contradict the principle exposed
in Section~\ref{sec:classical}, of the universality of Quantum Mechanics.
If everything in the universe is quantic and any two quantic states can be
superposed, then any two classical states, such as a live and a dead cat,
can be superposed. Many explanations have been proposed and this is not
the place for a survey. Most explanations accept the existence of 
superpositions of classical states and explain why such superpositions are
not seen. The analysis of the superposition concept to be developed below
proposes a radically different explanation. It is not the case, it is claimed 
here, that, in Quantum Mechanics, any two states can be superposed: 
on the contrary, no superposition of orthogonal states can ever be considered. 
Since different classical states are orthogonal,
the only superpositions of classical states that can ever occur are
trivial: superpositions of a state with itself. Trivial superpositions
are indeed observed and unproblematic.

To avoid any misunderstanding: if \mbox{$\mid + \rangle$} and 
\mbox{$\mid - \rangle$} are orthogonal states, the state 
\mbox{$1 / \sqrt{2} (\mid + \rangle + \mid - \rangle)$} is a perfectly legal
state, but it is not a superposition of \mbox{$\mid + \rangle$} and 
\mbox{$\mid - \rangle$}. It is equal, as will be clear, to many different
superpositions of non-orthogonal states (that are themselves linear 
combinations of the states \mbox{$\mid + \rangle$} and \mbox{$\mid - \rangle$}.
The reader will be well advised {\em not} to think {\em linear combination} 
when {\em superposition} is read.
 
To explain the surprising position above, let us, first, reflect on the 
nature of superpositions and their origin:
what are they and how do they come into consideration, without trying to
describe formally such superpositions. Then, we shall propose a formalization
and an algebraic structure.

The reader should notice that the linear combination of vectors of a Hilbert 
space provides a formal operation, not a conceptual analysis, and also that,
since vectors do not represent states, the linear combination of vectors
cannot offer a proper formalization for the superpositions of states.
Even though we announced above that orthogonal states cannot be superposed,
it is clear that orthogonal unit vectors can be combined linearly to form 
unit vectors. This should convince the reader that we shall not formalize
superposition as a straightforward linear combination.

\subsection{Nature and Origin} \label{sec:nature}
Superpositions must be considered to describe systems about which all we 
know is that they are the result of one of a number of different possible 
paths (or histories), 
i.e., if we have no way of knowing which history indeed took place.
In such a case, we must consider that the system is in some state that is
a superposition, i.e., a {\em compound} of the states that are the produced
by each of the possible paths. The term {\em compound} is used here
where, chemically-speaking, the term {\em mixture} may be more appropriate
because this last term is used in Quantum Mechanics with a different meaning.

If one knows which path has been taken, or one could discover 
which path has been taken, then one must consider that the system is in the
state that results from the path taken, and one must use probability theory
to describe one's ignorance about the state of the system.
If one does not know and cannot know which path has been taken, then one must
consider that the system is in some specific superposition of the states 
resulting from the different possible paths. This is a general principle:
if one cannot know which path has been taken, then those paths {\em interfere}
and therefore the system cannot be described using only probability theory,
but must be described by a state that is a compound, i.e., a superposition of
the states resulting from the different interfering paths.
This general principle holds also in Classical Physics, as will be seen in 
Section~\ref{sec:conditions}. The way in which the different paths may 
interfere, i.e., the parameters that characterize the different possible
superpositions will be described in Section~\ref{sec:parameters}.

The paradigmatic example of such a situation is a the two-slits
experiment in which a particle travels through one of two slits and
one does not know which.

\subsection{Parameters} \label{sec:parameters}
To leave things simple we shall consider only the superpositions of two
states, without loss of generality as long as we consider only a finite
number of possible paths. Generalizing to path integrals is beyond the scope
of this paper. Suppose therefore that we must deal with a system that may
result from two different paths. If path $p_{1}$ was taken, the system is 
in state $y$; if path $p_{2}$ was taken, the system is in state $z$.
If one cannot know which path was taken, one must consider that 
the system is in
a state that is some superposition of the two states $y$ and $z$. 
Many such superpositions are possible and the purpose of this section is
to describe the experimental parameters that influence the superposition
to be used.
In Section~\ref{sec:conditions}, the question of whether we can know which
path was taken will be given an unequivocal answer.

In a situation in which any one of two paths may have been taken, 
the experimental conditions determine the respective weights to be
given to each one of the possible paths. These relative weights may be 
interpreted as describing the a-priori probability of each one of the paths,
or the relative proportions in which each of the paths is taken.  
A superposition of $y$ and $z$ obtained
as the result of the interference between the two paths $p_{1}$ and $p_{2}$
will therefore be characterized by a single parameter \mbox{$r \in [0,1]$}. 
The proper value to be chosen
for this parameter is a function of the experimental setup. 
The reader should notice that, even though
we shall describe such a superposition of states $x$ and $y$ as some sort of 
{\em compound} or {\em mixture} of $x$ and $y$, a superposition is a pure
state, not what is known in QM as a mixed state.

The parameter $r$ that characterizes a superposition
describes, in a sense, the respective proportions (ratios) of $y$ and $z$
present in the superposition, though this intuitive analogy should not 
be taken too seriously. The parameter $r$ is therefore a real number:
\mbox{$0 \leq r \leq 1$} that describes the {\em weight} of $y$ relative to $z$
in the superposition.

In the two-slits experiment, where $y$ represents the state resulting from
the electron moving through the upper slit and $z$ the state resulting from
the electron moving through the lower slit, the parameter $r$ will depend on
the respective widths of the two slits and the respective distance of those
slits to the origin.

The superpositions we shall consider are therefore of the
form \mbox{$super(y, z, r)$} for states \mbox{$y, z \in X$} and
real number \mbox{$0 \leq r \leq 1$}.
The telling notation \mbox{$r y \, + \, (1 - r) z$} 
will be used in place of the more austere \mbox{$super(y, z, r)$},
but the reader is warned that $+$ does not mean addition, juxtaposition 
does not mean
multiplication and some of the properties one would expect from our notation
do {\em not} hold. In particular the composition of superpositions does not 
possess the properties suggested by the notation.

\subsection{Conditions} \label{sec:conditions}
Section~\ref{sec:parameters} indicated that superpositions of states $y$ and
$z$ should be considered only if there is no way to know which
one of the paths $p_{1}$ or $p_{2}$ leading to
$y$ and $z$ respectively has been traveled.
It is time to reflect on this condition.

If the states $y$ and $z$ are orthogonal: \mbox{$y \perp z$}, then there is
a way to find out for sure which of the two paths has been traveled:
perform on the resulting state a measurement testing whether the state is $y$
or not: a test $y$, $(\neg y)$. If path $p_{1}$ has been traveled, the result
will be a {\em yes} for sure since the state is $y$. If path $p_{2}$ 
has been traveled, the result, for sure, will be a {\em no} since the state
is $z$, orthogonal to $y$. Similarly, we could have tested for $z$ or for any
proposition satisfied by one of the states $y$ or $z$ and orthogonal to the
other one. We see that no superposition of orthogonal states can ever
be defined. This is is stark contrast with the linear combination of vectors
in a Hilbert space.

Further reflection shows that if the states $y$ and $z$ are not orthogonal,
one can never find out for sure which of the paths $p_{1}$ or $p_{2}$ has
been traveled. Indeed the only situation in which one could find out would be
to test for some proposition $\alpha$ satisfied, for sure, by one of the two
states $y$ or $z$ and not satisfied, for sure, by the other state.
In other terms, a closed subspace $\alpha$ containing one of $y$ or $z$ and
orthogonal to the other one. But this implies \mbox{$y \perp z$}.
We see that:
\[
{\bf Principle \ of \ Superposition} {\rm \ The \ superposition \ }
r y \, + \, (1 - r) z 
\]
\[ 
{\rm \ is \ defined \ if \ and \ only \ if \ } y \not \perp z.
\]

In Section~\ref{sec:superp_def} a definition of superpositions 
in the formalism of Hilbert spaces will be provided, 
but, first, we shall discuss two general principles, and justify them
by considerations independent of the Hilbert space formalism. 

\subsection{Trivial Superpositions} \label{sec:trivial}
Let us consider, first, the superpositions of a state $y$ with itself:
\mbox{$r y \, + \, (1 - r) y$}.
By the Principle of Classical Physics, these are the only superpositions
possible in classical physics.

Evidence from both classical and quantum physics shows that such superpositions
are trivial:
\[
{\bf Principle \ of \ Triviality} \ \forall y \in X, \forall r \in [0 , 1],
r y \, + \, (1 - r) y = y.
\]

Having disposed of the cases \mbox{$y \perp z$} and \mbox{$y = z$}, let us
study the generic case of superpositions.

\subsection{Principle of Coplanarity} \label{sec:coplanarity}

A superposition is coplanar with its components.
Assume \mbox{$y \not \perp z$}.
\[
{\bf Principle \ of \ Coplanarity} \  \forall r \in [0 , 1],
coplanar(r y \, + \, (1 - r) z, \, y, \, z) .
\]
This principle can be justified in the following way.
The superposition
\mbox{$x = r y \, + \, (1 - r) z$} results from
our inability to know which of $p_{1}$, resulting in $y$ or $p_{2}$,
resulting in $z$ has been traveled.
Measuring $\neg y$ on $x$ shows that the path $p_{1}$ has not been traveled
and therefore $p_{2}$ has been traveled and the current state
$(\neg y)(x)$ is in fact $(\neg y)(z)$.

We shall propose a precise definition of
superpositions such as \mbox{$r y \, + \, (1 - r) z$} for
\mbox{$y \not \perp z$} in Section~\ref{sec:superp_def}.
Then, in Sections~\ref{sec:euclidean} and~\ref{sec:theta}, 
we shall define fundamental geometric
quantities in terms of which the properties of superpositions 
will be studied in Section~\ref{sec:superp_prop}.

\section{Definition of Superpositions} \label{sec:superp_def}
We shall now present the definition of the superposition
\mbox{$r y \, + \, (1 - r) z$}.
Our definition is taken from the everyday practice of physicists.
\begin{definition} \label{def:superposition}
For any \mbox{$r \in [0 , 1]$}, 
for any \mbox{$y , z \in X$} such that \mbox{$y \not \perp z$},
we shall define
\mbox{$r y \, + \, (1 - r) z$} in the following way.

Choose some arbitrary unit vector $\vec{v}$ in $y$.
Since \mbox{$y \not \perp z$}, there is a unique unit vector $\vec{w}$
of $z$ such that \mbox{$\langle \vec{v} \, , \, \vec{w} \rangle \, > \, 0$}.
Define, now: 
\begin{equation} \label{eq:super}
\vec{u} \: = \: 
\sqrt{r} \, \vec{v} \, + \, \sqrt{1 - r} \, \vec{w}.
\end{equation}
Note that \mbox{$\vec{u} \neq \vec{0}$}: if \mbox{$y = z$} then 
\mbox{$\vec{v} = \vec{w}$} and \mbox{$\sqrt{r} + \sqrt{1 - r} > 0$}.
Otherwise $\vec{v}$ and $\vec{w}$ are
linearly independent and at least one of $\sqrt{r}$ or $\sqrt{1 - r}$ is
strictly positive.
We may now define \mbox{$r y \, + \, (1 - r) z$} 
to be the one-dimensional subspace generated by $\vec{u}$.
\end{definition}

Note that the vector $\vec{u}$ above is not a unit vector.
Definition~\ref{def:superposition} squares well with the Dirac notation
and the way it is used in everyday physics.
If $y$ and $z$ are to be compounded in equal parts (\mbox{$r = 1/2$}) then 
\mbox{$1/2 y + 1/2 z$} is defined by the vector 
\mbox{$1 / \sqrt{2} (\vec{v} + \vec{w})$}, which is a unit vector in case 
\mbox{$y \perp z$}. Notice, though, that the case $y$ and $w$ are orthogonal
is a case we do not allow.

The following is expected on general considerations and easily shown to
follow from Definition~\ref{def:superposition}.
\begin{lemma} \label{le:+commu}
For any \mbox{$y, z \in X$} such that \mbox{$y \not \perp z$}, we have
\begin{enumerate}
\item \label{one}
\mbox{$1 y \, + \, 0 z = y$}, and
\item \label{comm}
for any \mbox{$r \in [0 , 1]$}
\mbox{$r y \, + \, (1 - r) z =$}
\mbox{$(1 - r) z \, + \, r y$}.
\end{enumerate}
\end{lemma}

We shall now define two geometrical quantities that will help us understand
the structure of superpositions.

\section{The Geometry of Hilbert Spaces} \label{sec:geometry}
First, we shall define a geometrical property of two states.
\subsection{Quantities from Euclidean Geometry} \label{sec:euclidean}
\subsubsection{The Quantity $a(x, y)$} \label{sec:a}
We shall now define the first geometric quantity we wish to consider.
When considering the geometry of Hilbert spaces it is useful to
begin by reflecting on the geometry of Euclidean spaces, 
about which we know much more and have a much better intuition. 
Consider two lines, i.e, one-dimensional linear (not affine) subspaces, 
in $\cR^{n}$.
The only invariant characterizing their relation is their angle.
Two lines define a plane and four angles. Those four angles are two pairs
of equal angles. Therefore only two quantities are defined by two lines.
Moreover those two angles add up to $\pi$, therefore there is essentially
only one quantity defined. One can take as the fundamental quantity either
the acute or the obtuse angle. Let us consider the acute angle as the
quantity of interest. Two lines in Euclidean space define an angle $\varphi$
in the interval \mbox{$[0 , \pi / 2]$}. Equivalently, they define a real number
in the interval \mbox{$[0 , 1]$}, the value of $\cos(\varphi)$.

The same quantity may be defined in Hilbert spaces. 
Consider two states \mbox{$x, y \in X$}. We are trying to associate
a numerical quantity to this pair of states. The most natural thing to consider
is the inner product of two vectors contained in $x$ and $y$ respectively.
It is very natural to choose two unit vectors \mbox{$\vec{u} \in x$} and
\mbox{$\vec{v} \in y$} and consider the inner product 
\mbox{$\langle \vec{u} \, , \, \vec{v} \rangle$}. This will not do since the 
quantity depends on the choice of the unit vectors $\vec{u}$ and $\vec{v}$
and we are looking for a quantity that depends only on $x$ and $y$.
The inner product depends on the choice of the unit vectors, but its modulus
does not.
Consider therefore the quantity 
\[
a(x, y) \: \eqdef \: \mid \! \langle \vec{u} \, , \, \vec{v} \rangle \! \mid
\]
for arbitrary unit vectors $\vec{u}$ and $\vec{v}$ of $x$ and $y$ respectively.
Any unit vector $\vec{u}'$ of $x$ has the form:
\mbox{$\vec{u}' \, = \, e^{i \theta} \vec{u}$} and any $\vec{v}'$ of
$y$ has the form:
\mbox{$\vec{v}' \, = \, e^{i \varphi} \vec{v}$}.
Therefore
\mbox{$\langle \vec{u}' \, , \, \vec{v}' \rangle \, = \,$}
\mbox{$e^{i (\theta - \varphi)} \langle \vec{u} \, , \, \vec{v} \rangle$},
and 
\mbox{$\mid \langle \vec{u}' \, , \, \vec{v}' \rangle \mid \, = \,$}
\mbox{$\mid \langle \vec{u} \, , \, \vec{v} \rangle \mid$}.

The following is easily proved.
\begin{lemma} \label{le:a}
For any \mbox{$x, y \in X$}:
\begin{enumerate}
\item \label{r:01}
$a(x,y)$ is a real number of the interval $[0,1]$,
\item \label{r:one}
\mbox{$a(x,y) = 1$} iff \mbox{$x = y$},
\item \label{r:zero}
\mbox{$a(x,y) = 0$} iff \mbox{$x \perp y$},
\item \label{r:sym}
\mbox{$a(y,x) = a(x,y)$}.
\end{enumerate}
\end{lemma}

\subsubsection{Similarity: $p$} \label{sec:similarity}
It turns out that the square of the quantity \mbox{$a(x, y)$}, akin to the 
$\cos^2$ of an angle has even more remarkable properties.
\begin{definition} \label{def:p}
Given any states \mbox{$x, y \in X$}, we shall define their similarity
$p(x, y)$ by
\[
p(x, y) = a^{2}(x, y). 
\]
\end{definition}

The quantity $p$ will be called {\em similarity} because it measures
how similar, i.e., close, are its arguments $x$ and $y$. Its physical 
interpretation is straightforward: $p(x,y)$ is the probability that,
when, on state $x$, one tests whether $y$ is the case, one gets a positive 
answer. With probability $1 - p(x,y)$ one gets the the answer that $y$ is
not the case. This physical interpretation is the reason $p = a^{2}$ 
and not $a$ has been chosen as the quantity of reference. 
Note that $p$ can be directly obtained experimentally.
Below, we shall extend the definition of $p$ to measure
the {\em similarity} between any state \mbox{$x \in X$} and any proposition
\mbox{$\alpha \in M$}, i.e., the degree to which state $x$ satisfies 
proposition $\alpha$.

A straightforward result on Hilbert spaces will be recalled now.
\begin{lemma} \label{le:inner}
Let \mbox{$\vec{u}, \vec{v} \in \cH$}.
Assume $\vec{v}$ is a unit vector and \mbox{$\vec{v} \in x \in X$}.
Then the projection \mbox{$x(\vec{u})$} of $\vec{u}$ on $x$ is 
\mbox{$\langle \vec{u} \, , \, \vec{v} \rangle \, \vec{v}$}.
\end{lemma}
\begin{proof}
\mbox{$\vec{u} - \langle \vec{u} \, , \, \vec{v} \rangle \, \vec{v}$} 
is indeed 
orthogonal to $\vec{v}$ and therefore to $x$.
\end{proof}

First properties of $p$ are described in the following.
\begin{lemma} \label{le:p}
For any \mbox{$x, y \in X$}:
\begin{enumerate}
\item \label{p:01}
$p(x,y)$ is a real number in the interval $[0,1]$,
\item \label{p:one}
\mbox{$p(x,y) = 1$} iff \mbox{$x = y$},
\item \label{p:zero}
\mbox{$p(x,y) = 0$} iff \mbox{$x \perp y$},
\item \label{p:sym}
\mbox{$p(y,x) = p(x,y)$},
\item \label{p:inner}
for any unit vector \mbox{$\vec{u} \in x$},
\mbox{$p(x,y) \, = \,$}
\mbox{$\langle \vec{u} \, , \, y(\vec{u}) \rangle$} 
where $y(\vec{u})$ is the projection of 
$\vec{u}$ on $y$,
\item \label{p:proj}
for any unit vector \mbox{$\vec{u} \in x$},
\mbox{$p(x,y) \, = \,$}
\mbox{$\parallel \! y(\vec{u}) \! \parallel^{2}$}.
\end{enumerate}
\end{lemma}
\begin{proof}
For~\ref{p:inner}, note that 
for any unit vector $\vec{v}$ of $y$, we have, by Lemma~\ref{le:inner},
\mbox{$y(\vec{u}) \, = \, \langle \vec{u} \, , \, \vec{v} \rangle \, \vec{v}$},
and therefore
\mbox{$\langle \vec{u} \, , \, y(\vec{u}) \rangle \, = \,$}
\mbox{$\overline{\langle \vec{u} \, , \, \vec{v} \rangle} \,  
\langle \vec{u} \, , \, \vec{v} \rangle \, = \,$}
\mbox{$\mid \! \langle \vec{u} \, , \, \vec{v} \rangle \! \mid^{2}$}.
Note that this implies that the inner product
\mbox{$\langle \vec{u} \, , \, y(\vec{u}) \rangle$} is a real number.
For~\ref{p:proj}, note that projections are Hermitian and idempotent, and
therefore
\mbox{$\langle y(\vec{u}) \, , \, y(\vec{u}) \rangle \, = \,$}
\mbox{$\langle \vec{u} \, , \, y(y(\vec{u})) \rangle \, = \,$}
\mbox{$\langle \vec{u} \, , \, y(\vec{u}) \rangle$}.
\end{proof}

The next result is central. 
It shows that, for any given proposition $\alpha$, the projection
on $\alpha$ is determined by the $p$-structure.
\begin{theorem} \label{the:p}
For any proposition \mbox{$\alpha \in M$} and any
states \mbox{$x, y \in X$}, if \mbox{$x \not \perp \alpha$}
and \mbox{$y \in \alpha$} then
\mbox{$p(x,y) = p(x,\alpha(x)) \, p(\alpha(x), y)$}. 
\end{theorem}
\begin{proof}
Let $\vec{u}$ be a unit vector of $x$.
Since \mbox{$y \in \alpha$}, the projection of any vector on $y$
can be obtained by projecting the vector first on $\alpha$ and then
projecting the result on $y$. In particular,
\mbox{$y(\vec{u}) = y(\alpha(\vec{u}))$}.
Therefore
\[
p(x,y) = \parallel \! y(\vec{u}) \! \parallel^{2} = 
\parallel \! y(\alpha(\vec{u})) \! \parallel^{2} / 
\parallel \! \alpha(\vec{u}))\! \parallel^{2} /: \times \:
\parallel \! \alpha(\vec{u}) \! \parallel^{2}
\]
Let \mbox{$\vec{v} =$}
\mbox{$ \alpha(\vec{u}) / \parallel \! \alpha(\vec{u} \! \parallel$}. 
Notice that $\vec{v}$ is a unit vector of $\alpha(x)$ and therefore
\[
p(x,y) = \parallel \! \vec{v} \! \parallel^{2} \: \times \:
\parallel \! \alpha(\vec{u}) \! \parallel^{2} = 
p(\alpha(x), y) \: \times \: p(x, \alpha(x))
\]
since $\alpha(\vec{u})$ is the projection of $\vec{u}$ on $\alpha(x)$,
and by Lemma~\ref{le:p}.
\end{proof}

\begin{corollary} \label{le:max}
For any proposition \mbox{$\alpha \in M$} and any
state \mbox{$x \in X$}, if \mbox{$x \not \perp \alpha$}
then $\alpha(x)$ is the unique state $y$ of $\alpha$ on which
the value of $p(x,y)$ is maximal. 
\end{corollary}
In short, there is a unique state of $\alpha$ that is most similar to $x$,
this is $x$'s projection on $\alpha$.
\begin{proof}
By Theorem~\ref{the:p}, since \mbox{$p(\alpha(x),y) \leq 1$} by
Lemma~\ref{le:p}, \mbox{$p(x,y) \leq p(x,\alpha(x))$} for any
\mbox{$y \in \alpha$}.

For uniqueness, suppose \mbox{$y \in \alpha$} and 
\mbox{$p(x,y) = p(x,\alpha(x))$}.
By Theorem~\ref{the:p}, 
\mbox{$p(x,\alpha(x)) =$}
\mbox{$p(x, \alpha(x)) \, p(\alpha(x), y)$}.
Since $x$ is not orthogonal to $\alpha$, \mbox{$p(x,\alpha(x)) > 0$}
and therefore \mbox{$p(\alpha(x),y) = 1$} and \mbox{$\alpha(x) = y$}.
\end{proof}

It is now only natural to extend the definition of $p$ to an arbitrary
proposition as second argument. 
For any \mbox{$x \in X$} and 
\mbox{$\alpha \in M$}, we define \mbox{$p(x,\alpha)$} in the following way:
\begin{itemize}
\item \mbox{$p(x,\alpha) = 0$} if \mbox{$x \perp \alpha$}, and
\item \mbox{$p(x,\alpha) = p(x,\alpha(x))$} otherwise.
\end{itemize}

The following is known, in Physics, as Born's rule.
The quantity $p(x,\alpha)$ is the probability of measuring the property
$\alpha$ in state $x$.
\begin{lemma} \label{le:Born}
For any state $x \in X$ and any proposition \mbox{$\alpha \in M$}, if
\mbox{$\vec{u} \neq \vec{0} \in x$},
\mbox{$p(x, \alpha) = \parallel \alpha(\vec{u}) \parallel^{2} / 
\parallel \vec{u} \parallel^{2}$}.
\end{lemma}
The proof is obvious. The following is an obvious consequence of 
Corollary~\ref{le:max}.
\begin{corollary} \label{le:satisfaction}
For any state $x$ and any proposition $\alpha$,
\mbox{$x \in \alpha$} iff \mbox{$\alpha(x) = x$} iff 
\mbox{$p(x, \alpha) = 1$}.
\end{corollary}

The next two sections prove additional properties of the quantity $p$.
On a first reading the reader is advised to advance to 
Section~\ref{sec:theta}. 
Section~\ref{sec:probas} shows that, 
for any given $x$ and different $\alpha$'s, \mbox{$p(x, \alpha)$} 
behaves very much as a probability on the propositions. 
Exactly so, for propositions that commute as projections.
Section~\ref{sec:num_inter} proves an intriguing inequality that
provides a numerical strengthening of the {\bf Interference} property
of~\cite{LEG:Malg}.

\subsubsection{Similarity as Probability} \label{sec:probas}
The following results will show that, for any fixed \mbox{$x \in X$}, 
the quantities
$p(x, \alpha)$ for different measurements $\alpha$ play the role of a 
probability on the propositions.
For any two propositions \mbox{$\alpha, \beta \in M$} we shall define, as
traditional since~\cite{BirkvonNeu:36},
their conjunction \mbox{$\alpha \wedge \beta$} as their intersection
\mbox{$\alpha \cap \beta$} (note the intersection of closed subspaces
is a closed subspace) and their disjunction \mbox{$\alpha \vee \beta$}
as the topological closure of their linear sum: 
\mbox{$cl(\alpha + \beta)$}.
Note that these notations are inconsistent with those of~\cite{LEG:Malg}
where conjunction and disjunction were defined only for {\em commuting}
propositions.
We shall demonstrate a particular interest in {\em commuting} propositions.
For the sake of obtaining a straightforward definition of commutation, 
we shall extend our notation for projections.
\begin{definition} \label{def:commuting}
Let \mbox{$\alpha, \beta \in M$} be two propositions. We shall say that
$\alpha$ and $\beta$ {\em commute} iff for any \mbox{$x \in X$} 
\mbox{$\alpha(\beta(x)) = \beta(\alpha(x))$}.
\end{definition}

\begin{lemma} \label{le:commuting}
Any two propositions \mbox{$\alpha, \beta \in M$} commute iff
there are three pairwise orthogonal propositions 
\mbox{$\gamma_{i}, i = 1, \ldots , 3$} such that
\mbox{$\alpha =$} \mbox{$\gamma_{1} \vee \gamma_{2}$} and
\mbox{$\beta =$} \mbox{$\gamma_{1} \vee \gamma_{3}$}.
\end{lemma}
Note that one of the propositions $\gamma_{i}$ may be the falsehood $\bot$.
\begin{proof}
The {\em if} claim is obvious. The {\em only if} claim follows 
from the fact that 
projections are Hermitian and that Hermitian operators commute iff they
have a joint basis of eigenvectors.
\end{proof}

\begin{corollary} \label{le:comm}
For any \mbox{$\alpha, \beta \in X$}, if \mbox{$\alpha \subseteq \beta$}
or \mbox{$\alpha \perp \beta$}, then $\alpha$ and $\beta$ commute.
\end{corollary}
\begin{proof}
In the first case, take \mbox{$\gamma_{1} = \alpha$}, 
\mbox{$\gamma_{2} = \bot$} and \mbox{$\gamma_{3} =$}
\mbox{$\neg \alpha \wedge \beta$}.
In the second case, take \mbox{$\gamma_{1} = \alpha$}, 
\mbox{$\gamma_{2} = \bot$} and \mbox{$\gamma_{3} = \beta$}.
\end{proof}

\begin{corollary} \label{le:comm_neg}
For any \mbox{$\alpha, \beta \in X$}, if $\alpha$ and $\beta$ commute
then $\neg \alpha$ and $\beta$ commute.
\end{corollary}
\begin{proof}
Let \mbox{$\alpha =$} \mbox{$\gamma_{1} \vee \gamma_{2}$} and
\mbox{$\beta =$} \mbox{$\gamma_{1} \vee \gamma_{3}$}.
Then \mbox{$\neg \alpha =$} \mbox{$\neg \gamma_{1} \wedge \neg \gamma_{2}$}.
Since \mbox{$\gamma_{3} \subseteq \neg \alpha$}, we have, by the
orthomodular property,
\mbox{$\neg \alpha =$} 
\mbox{$\gamma_{3} \vee \neg \gamma_{1} \wedge \neg \gamma_{2} 
\wedge \neg \gamma_{3}$}.
But \mbox{$\beta =$} \mbox{$\gamma_{3} \vee \gamma_{2}$} and
\mbox{$\gamma_{2} \perp \neg \gamma_{1} \wedge \neg \gamma_{2}
\wedge \neg \gamma_{3}$}.
\end{proof}

First, we shall consider disjunctions of orthogonal propositions.
\begin{lemma} \label{le:orthodisjunction}
If \mbox{$\alpha \perp \beta$} then, for any \mbox{$x \in X$},
\mbox{$p(x, \alpha \vee \beta) =$}
\mbox{$p(x, \alpha) + p(x, \beta)$}.
\end{lemma}
\begin{proof}
Consider any \mbox{$\vec{u} \neq \vec{0} \in x$}.
Now \mbox{$(\alpha \vee \beta)(\vec{u}) =$}
\mbox{$\alpha(\vec{u}) + \beta(\vec{u})$} (see~\cite{Halmos:Hilbert} 
Theorem 2, page 46). Therefore
\mbox{$\langle \vec{u} \, , \, (\alpha \vee \beta)(\vec{u}) \rangle =$}
\mbox{$\langle \vec{u} \, , \, \alpha(\vec{u}) \rangle +$}
\mbox{$\langle \vec{u} \, , \, \beta(\vec{u}) \rangle$}. 
\end{proof}
\begin{corollary} \label{co:orthodisjunction}
If \mbox{$\alpha_{i}$} is a family of pairwise orthogonal measurements, then
for any \mbox{$x \in X$} we have
\mbox{$p(x, \bigvee_{i \in I} \alpha_{i}) =$}
\mbox{$\sum_{i \in I} p(x, \alpha_{i})$}.
\end{corollary}
\begin{proof}
By induction on the size of $I$, and associativity of disjunction.
\end{proof}

The following lemmas are fundamental characteristics of probabilities.
\begin{lemma} \label{le:neg_sum_one}
For any \mbox{$\alpha \in M$} and any \mbox{$x \in X$}:
\mbox{$p(x, \alpha) + p(x, \neg \alpha) = 1$}.
\end{lemma}
\begin{proof}
By Lemma~\ref{le:orthodisjunction}, 
\mbox{$p(x, \alpha) + p(x, \neg \alpha) = p(x, \alpha \vee \neg \alpha)$}.
But \mbox{$\alpha \vee \neg \alpha = \top$} and therefore 
\mbox{$(\alpha \vee \neg \alpha)(x) = x$} and, 
by Corollary~\ref{le:satisfaction},
\mbox{$p(x, \alpha \vee \beta) = 1$}.
\end{proof}
\begin{lemma} \label{le:zero_one}
For any \mbox{$\alpha \in M$} and any \mbox{$x \in X$}:
\mbox{$0 \leq p(x, \alpha) \leq 1$}.
\end{lemma}
\begin{proof}
By Lemmas~\ref{le:Born} and~\ref{le:neg_sum_one}.
\end{proof}
\begin{lemma} \label{le:disj_prob}
Let \mbox{$\alpha, \beta \in M$} be any {\em commuting} measurements.
For any \mbox{$x \in X$}
\mbox{$p(x, \alpha \vee \beta) =$}
\mbox{$p(x, \alpha) + p(x, \beta) - p(x, \alpha \wedge \beta)$}.
\end{lemma}
\begin{proof}
We know that \mbox{$\alpha \vee \beta =$}
\mbox{$(\alpha \wedge \beta) \vee (\alpha \wedge \neg \beta)
\vee (\neg \alpha \wedge \beta)$}. The three parts of the disjunction above
are pairwise orthogonal, therefore Corollary~\ref{co:orthodisjunction} implies
that \mbox{$p(x, \alpha \vee \beta) =$}
\mbox{$p(x, \alpha \wedge \beta) +$}
\mbox{$p(x, \alpha \wedge \neg \beta) +$}
\mbox{$p(x, \neg \alpha \wedge \beta)$}.
But, by Lemma~\ref{le:orthodisjunction}:
\mbox{$p(x, \alpha \wedge \beta) +$}
\mbox{$p(x, \alpha \wedge \neg \beta) =$}
\mbox{$p(x, \alpha)$} and
\mbox{$p(x, \alpha \wedge \beta) +$}
\mbox{$p(x, \neg \alpha \wedge \beta) =$}
\mbox{$p(x, \beta)$}.
\end{proof}
The lemmas above dealt mostly with the properties of disjunction. The next
result concerns conjunction and parallels the consideration of conditional
probabilities.
\begin{lemma} \label{le:conj}
Let \mbox{$\alpha, \beta \in M$} be any {\em commuting} measurements.
For any \mbox{$x \in X$}:
\mbox{$p(x, \alpha \wedge \beta) =$}
\mbox{$p(x, \alpha) \: p(\alpha(x), \beta)$}.
\end{lemma}
\begin{proof}
Since \mbox{$\alpha \wedge \beta = \alpha \circ \beta$}, by the definition
of $p$, taking any \mbox{$\vec{u} \neq \vec{0} \in x$}:
\[
p(x, \alpha \wedge \beta) =
{{\parallel (\alpha \circ \beta)(\vec{u}) \parallel^{2}} \over
{\parallel \vec{u} \parallel^{2}}} =
{{\parallel (\alpha \circ \beta)(\vec{u}) \parallel^{2}} \over
{\parallel \alpha(\vec{u}) \parallel^{2}}} \ 
{{\parallel \alpha(\vec{u}) \parallel^{2}} \over 
{\parallel \vec{u} \parallel^{2}}} =
p(\alpha(x), \beta) \  p(x, \alpha).
\]
\end{proof}
\begin{corollary} \label{co:leq}
Let \mbox{$\alpha, \beta \in M$} be any measurements such that 
\mbox{$\alpha \leq \beta$}.
Then for any \mbox{$x \in X$}, \mbox{$p(x, \alpha) \leq p(x, \beta)$}.
\end{corollary}
\begin{proof}
If \mbox{$\alpha \leq \beta$}, the two measurements commute and
\mbox{$\alpha = \beta \wedge \alpha$}.
By Lemma~\ref{le:conj}, then \mbox{$p(x, \alpha) =$}
\mbox{$p(x, \beta) \: p(\beta(x), \alpha) \leq$}
\mbox{$p(x, \beta)$} by Lemma~\ref{le:zero_one}. 
\end{proof}
\begin{corollary} \label{co:comp}
Let \mbox{$\alpha, \beta \in M$} be any {\em commuting} measurements.
Then for any \mbox{$x \in X$}, 
\mbox{$p(x, \beta) = p(x, \alpha) \, p(\alpha(x), \beta) \: + \: 
p(x, \neg \alpha) \, p((\neg \alpha)(x), \beta)$}.
\end{corollary}
\begin{proof}
Since $\alpha$ and $\beta$ commute, by Theorem~1 of~\cite{LEG:Malg},
\mbox{$\beta = (\alpha \wedge \beta) \vee (\neg \alpha \wedge \beta)$}.
By Lemma~\ref{le:orthodisjunction} we have:
\mbox{$p(x, \beta) = p(x, \alpha \wedge \beta) \: + \: p(x, \neg \alpha \wedge \beta)$}.
We conclude, by Lemma~\ref{le:conj}, that
\mbox{$p(x, \beta) = p(x, \alpha) \, p(\alpha(x), \beta) \: + \: 
p(x, \neg \alpha) \, p((\neg \alpha)(x), \beta)$}.
\end{proof}
In Corollary~\ref{co:comp} one cannot omit the requirement 
that $\alpha$ and $\beta$ commute. The consideration of a two-dimensional 
Euclidean space where $\alpha$ is the x-axis and $x$ makes an angle
$\theta$ with the x-axis is sufficient. If $\beta$ is $x$, then
\mbox{$p(x, \beta) = 1$} whereas 
\mbox{$p(x, \alpha) =$}
\mbox{$\cos^{2}(\theta) =$} \mbox{$p(\alpha(x), \beta)$} and
\mbox{$p(x, \neg \alpha) =$}
\mbox{$\sin^{2}(\theta) =$} \mbox{$p((\neg \alpha)(x), \beta)$}.
Also taking $\beta$ orthogonal to $x$ gives
\mbox{$p(x, \beta) = 0$} and 
\mbox{$p(x, \alpha) =$} \mbox{$\cos^{2}(\theta) =$}
\mbox{$p((\neg \alpha)(x), \beta)$} and
\mbox{$p(x, \neg \alpha) =$} \mbox{$\sin^{2}(\theta) =$}
\mbox{$p(\alpha(x), \beta)$}.
Nevertheless the result holds in the following case.
\begin{lemma} \label{le:orthomodular_equality}
For any \mbox{$x \in X$} and any \mbox{$\alpha, \beta \in M$} such that
\mbox{$\alpha(x) \in \beta$} and \mbox{$(\neg \alpha)(x) \in \beta$},
one has 
\[
p(x, \beta) = p(x, \alpha) \, p(\alpha(x), \beta) \: + \: 
p(x, \neg \alpha) \, p((\neg \alpha)(x), \beta)= 1.
\]
\end{lemma}
\begin{proof}
By assumption both $\alpha(x)$ and $(\neg \alpha)(x)$ are subspaces of 
$\beta$. Given any \mbox{$\vec{u} \in x$}, both 
$\alpha(\vec{u})$ and $(\neg \alpha)(\vec{u})$ are in $\beta$.
But $\beta$ is a subspace and therefore 
\mbox{$\alpha(\vec{u}) + (\neg \alpha)(\vec{u}) =$}
\mbox{$\vec{u} \in \beta$}.
\end{proof}

\begin{lemma} \label{le:local_comp}
For any \mbox{$x \in X$} and any \mbox{$\alpha, \beta \in M$} such that
\mbox{$(\alpha \circ \beta)(x) =$} \mbox{$(\beta \circ \alpha)(x)$}, we have
\mbox{$p(x, \beta) =$} \mbox{$p(x, \alpha) \, p(\alpha(x), \beta) \: + \: 
p(x, \neg \alpha) \, p((\neg \alpha)(x), \beta)$}.
\end{lemma}
\begin{proof}
Assume that \mbox{$(\alpha \circ \beta)(x) =$}
\mbox{$(\beta \circ \alpha)(x)$}.
By Lemma~\ref{le:comm_neg}, 
\mbox{$(\neg \alpha \circ \beta)(x) =$} 
\mbox{$(\beta \circ \neg \alpha)(x)$}.
Take any \mbox{$\vec{u} \neq \vec{0} \in x$}.
Then,
\[
p(x, \beta) \ = \  
\parallel \beta(\vec{u}) \parallel^{2} \: / \: 
\parallel \vec{u} \parallel^{2} \ = \ 
\parallel \alpha(\beta(\vec{u}))) + (\neg \alpha)(\beta(\vec{u})) \parallel^{2}
\: / \: \parallel \vec{u} \parallel^{2} \ = 
\]
\[
\parallel \alpha(\beta(\vec{u}))) \parallel^{2}\: / \: 
\parallel \vec{u} \parallel^{2}  
+ \parallel (\neg \alpha)(\beta(\vec{u})) \parallel^{2}
\: / \: \parallel \vec{u} \parallel^{2} \ = \ 
\]
\[
\parallel \beta(\alpha(\vec{u}))) \parallel^{2}\: / \: 
\parallel \vec{u} \parallel^{2}  
+ \parallel (\beta)((\neg \alpha)(\vec{u})) \parallel^{2}
\: / \: \parallel \vec{u} \parallel^{2} \ = \ 
\]
\[
{{\parallel \beta(\alpha(\vec{u})) \parallel^{2}} \over
{\parallel \alpha(x) \parallel^{2}}} \ 
{{\parallel \alpha(x) \parallel^{2}} \over
{\parallel \vec{u} \parallel^{2}}}
+ 
{{\parallel (\beta)((\neg \alpha)(\vec{u})) \parallel^{2}} \over 
{\parallel (\neg \alpha)(x) \parallel^{2}}} \ 
{{\parallel (\neg \alpha)(x) \parallel^{2}} \over 
{\parallel \vec{u} \parallel^{2}}} \ = \ 
\]
\[
p(\alpha(x),\beta) \, p(x, \alpha) \: + \: p((\neg \alpha)(x), \beta) \, 
p(x, \neg \alpha). 
\]
\end{proof}

\subsubsection{An Inequality} \label{sec:num_inter}
The next result strengthens the Interference property of~\cite{LEG:Malg}
by presenting a quantitative version of the principle.
\begin{theorem} \label{the:quant_interference}
For any \mbox{$\alpha, \beta \in M$} and any \mbox{$x \in X$} such that
\mbox{$\alpha(x) = x$}, 
\[
p(x, \beta) \: (1 - p(\beta(x), \alpha))^{2} \: \leq \:
p(\beta(x), \alpha) \: (1 - p(\alpha(\beta(x)), \beta))
\]
\end{theorem}
Note that, by Theorem~\ref{the:p},
\mbox{$p(x, \beta) \leq p(\beta(x), \alpha)$} but
\mbox{$(1 - p(\beta(x), \alpha)) \geq (1 - p(\beta(x), \alpha))$}.
The fact that the quantity \mbox{$1 - p(\beta(x), \alpha)$} 
appears squared seems inevitable. 
An examination of $\cR^{3}$ shows that it may be the case that
\mbox{$p(x, \beta) \: (1 - p(\beta(x), \alpha)) \: > \:$}
\mbox{$p(\beta(x), \alpha) (1 - p(\alpha(\beta(x)), \beta))$}.
\begin{proof}
Assume \mbox{$\vec{t} \neq \vec{0} \in x$}.
Let \mbox{$\vec{u} = \beta(\vec{t})$}, \mbox{$\vec{v} = \alpha(\vec{u})$}
and \mbox{$\vec{w} = \beta(\vec{v})$}.

In a first step we want to show that:
\[
\parallel \vec{u} - \vec{v} \parallel^{2} =
\langle \vec{t} \, , \, \vec{v} - \vec{w} \rangle.
\]

Indeed: \mbox{$\parallel \vec{u} - \vec{v} \parallel^{2} =$}
\mbox{$\langle \vec{u} - \vec{v} \, , \, \vec{u} - \vec{v} \rangle =$}
\mbox{$\langle \vec{u} \, , \, \vec{u} - \vec{v} \rangle -$}
\mbox{$\langle \vec{v} \, , \, \vec{u} - \vec{v} \rangle$}.
But the last term is null since \mbox{$\vec{u} - \vec{v}$} is orthogonal
to $\alpha$ in general and in particular to $\vec{v}$.
We have:
\[
\parallel \vec{u} - \vec{v} \parallel^{2} =
\langle \vec{u} \, , \, \vec{u} - \vec{v} \rangle.
\]
But \mbox{$\vec{t} - \vec{u}$} is, similarly, orthogonal
to $\vec{u}$ and \mbox{$\langle \vec{u} \, , \, \vec{u} \rangle =$}
\mbox{$\langle \vec{t} \, , \, \vec{u} \rangle$}.
Since \mbox{$\vec{u} - \vec{v}$} is orthogonal to $\vec{t}$,
\mbox{$\langle \vec{t} \, , \, \vec{u} \rangle =$}
\mbox{$\langle \vec{t} \, , \, \vec{v} \rangle$}.
We have:
\[
\parallel \vec{u} - \vec{v} \parallel^{2} =
\langle \vec{t} \, , \, \vec{v} \rangle - \langle \vec{u} \, , \, \vec{v} \rangle.
\]
Again, \mbox{$\vec{v} - \vec{w}$} is orthogonal to $\vec{u}$ and therefore:
\mbox{$\langle \vec{u} \, , \, \vec{v} \rangle =$}
\mbox{$\langle \vec{u} \, , \, \vec{w} \rangle$} 
and \mbox{$\vec{t} - \vec{u}$} is orthogonal to $\vec{w}$ and
we have: \mbox{$\langle \vec{u} \, , \, \vec{w} \rangle =$}
\mbox{$\langle \vec{t} \, , \, \vec{w} \rangle$}.
Therefore:
\[
\parallel \vec{u} - \vec{v} \parallel^{2} =
\langle \vec{t} \, , \, \vec{v} \rangle - \langle \vec{t} \, , \, \vec{w} \rangle =
\langle \vec{t} \, , \, \vec{v} - \vec{w} \rangle.
\]
By Cauchy-Schwarz therefore we have:
\[
\parallel \vec{u} - \vec{v} \parallel^{2} \: \leq \:
\parallel \vec{t} \parallel \: \parallel \vec{v} - \vec{w} \parallel.
\]
and:
\[
\parallel \vec{u} - \vec{v} \parallel^{4} \: \leq \:
\parallel \vec{t} \parallel^{2} \: \parallel \vec{v} - \vec{w} \parallel^{2}.
\]
But: \mbox{$\parallel \vec{u} \parallel^{2} =$}
\mbox{$\parallel \vec{v} \parallel^{2} +$}
\mbox{$\parallel \vec{u} - \vec{v} \parallel^{2}$}, and
\mbox{$\parallel \vec{v} \parallel^{2} =$}
\mbox{$\parallel \vec{w} \parallel^{2} +$}
\mbox{$\parallel \vec{v} - \vec{w} \parallel^{2}$}.
Therefore we have:
\[
(\parallel \vec{u} \parallel^{2} - \parallel \vec{v} \parallel^{2})^{2} 
\: \leq \:
\parallel \vec{t} \parallel^{2} \: 
(\parallel \vec{v} \parallel^{2} - \parallel \vec{w} \parallel^{2}).
\]
and
\[
{{\parallel {\vec{u}} \parallel^{2}} \over 
{\parallel \vec{t} \parallel^{2}}}
\: ( 1 - 
{{\parallel \vec{v} \parallel^{2}} \over 
{\parallel \vec{u} \parallel^{2}}})^{2}
\: \leq \:
{{\parallel \vec{v} \parallel^{2} - \parallel \vec{w} \parallel^{2}}
\over
{\parallel \vec{u} \parallel^{2}}},
\]
\[
p(x, \beta) \: ( 1 - p(\beta(x), \alpha))^{2} \: \leq \:
{{\parallel \vec{v} \parallel^{2}} \over {\parallel \vec{u} \parallel^{2}}}
\: (1 - {{\parallel \vec{w} \parallel^{2}} \over 
{\parallel \vec{v} \parallel^{2}}}).
\]
We conclude that:
\[
p(x, \beta) \: ( 1 - p(\beta(x), \alpha))^{2} \: \leq \:
p(\beta(x), \alpha) \: (1 - p(\alpha(\beta(x)), \beta)).
\]
\end{proof}
Theorem~\ref{the:quant_interference} is a quantitative strengthening of the
{\bf Interference} property of projections in Hilbert spaces that plays a 
central role in the definition of an M-algebra~\cite{LEG:Malg}.
Indeed, assuming that \mbox{$x \in \alpha$}, if 
\mbox{$\alpha(\beta(x)) \in \beta$}, then, by Corollary~\ref{le:satisfaction},
\mbox{$p(\alpha(\beta(x)), \beta) = 1$} and by 
Theorem~\ref{the:quant_interference}, either \mbox{$p(x, \beta) = 0$}
or \mbox{$p(\beta(x), \alpha) = 1$}. In both cases we have 
\mbox{$p(\beta(x), \alpha) = 1$} and, by  Corollary~\ref{le:satisfaction},
\mbox{$\beta(x) \in \alpha$}.

\subsection{Phases for Triangles: $\theta(x, y, z)$} 
\label{sec:theta}
We may now proceed to the definition of a second geometric quantity 
relating three states: $\theta(x, y, z)$. This quantity does not seem to
have been studied previously.

In section~\ref{sec:a} a quantity was attached to any pair of states.
This quantity was the modulus of some inner product.
It seems natural that the argument of a similar inner product represents
another important geometrical quantity. But, clearly some thinking must be
done to define, out of such an argument, a quantity that does not depend
on the vectors chosen, but only on states.
A new quantity, \mbox{$\theta(x, y, z)$}, 
an angle in the interval $[0, 2 \pi]$
will be attached to triples of states.
This quantity can be defined only if no two of the three states
$x$, $y$ and $z$ are orthogonal.

\begin{definition} \label{def:theta}
Let \mbox{$x, y, z \in X$} be such that \mbox{$x \not \perp y$},
\mbox{$y \not \perp z$} and \mbox{$z \not \perp x$}.
We shall define
\mbox{$\theta(x, y, z)$} in the following way.
Choose arbitrary unit vectors
$\vec{u}$, $\vec{v}$ and $\vec{w}$ in $x$, $y$ and $z$ respectively 
and let:
\[
\theta(x, y, z) \: = \: \arg(\langle \vec{u} \, , \, \vec{v} \rangle) \: + \:
\arg(\langle \vec{v} \, , \, \vec{w} \rangle) \: + \:
\arg(\langle \vec{w} \, , \, \vec{u} \rangle).
\]

Note that each of those three inner products is different from zero, 
by assumption, and therefore the three complex arguments are well-defined. 
\end{definition}
We need to justify the definition by showing that the quantity 
$\theta(x, y, z)$ depends only on $x$, $y$ and $z$ and does not
depend on the vectors $\vec{u}$, $\vec{v}$ and $\vec{w}$.
For example, the definition is independent of the vector $\vec{u}$ 
chosen in $x$ since any unit vector $\vec{s}$ of $x$ has the form
\mbox{$\vec{s} = e^{i \varphi} \vec{u}$} for some
\mbox{$\varphi \in [0 , 2 \pi]$}.
Had we used $\vec{s}$ instead of $\vec{u}$ we would have obtained:
\[
\arg(\langle e^{i \varphi} \vec{u} \, , \, \vec{v} \rangle) \: + \:
\arg(\langle \vec{v} \, , \, \vec{w} \rangle) \: + \:
\arg(\langle \vec{w} \, , \, e^{i \varphi} \vec{u} \rangle) \: = \:
\]
\[
\arg(e^{i \varphi} \langle \vec{u} \, , \, \vec{v} \rangle) \: + \:
\arg(\langle \vec{v} \, , \, \vec{w} \rangle) \: + \:
\arg(e^{- i \varphi} \langle \vec{w} \, , \, \vec{u} \rangle) \: = \:
\]
\[
\varphi + \arg(\langle \vec{u} \, , \, \vec{v} \rangle) \: + \:
\arg(\langle \vec{v} \, , \, \vec{w} \rangle) \: - \: \varphi \: + \:
\arg(\langle \vec{w} \, , \, \vec{u} \rangle).
\]
A similar line shows that the choice of none of $\vec{v}$ or
$\vec{w}$ influences $\theta(x, y, z)$.

We shall now prove some properties of $\theta$.
First, \mbox{$\theta(x, y, z)$} is invariant under a circular permutation
of the arguments and antisymmetric under transpositions.
\begin{lemma} \label{le:cyclic}
For any generic states $x$, $y$ and $z$, we have:
\mbox{$\theta(y, z, x) =$} \mbox{$\theta(x, y, z)$},
\mbox{$\theta(x, z, y) =$} \mbox{$- \theta(x, y, z)$} and
\mbox{$\theta(x, y, w) =$}
\mbox{$\theta(x, y, z) + \theta(x, z, w) + \theta(z, y, w)$}.
\end{lemma}
\begin{proof}
Obvious.
\end{proof}
The behavior of $\theta$ under (planar) orthogonal complements is also
antisymmetric.
\begin{lemma} \label{le:prime}
Assume \mbox{$x, y, z \in X$} are states no two of them are equal
and no two of them are orthogonal
and such that \mbox{$coplanar(x, y, z)$}.
Let \mbox{$x' =$} \mbox{$(\neg x)(y) =$} \mbox{$(\neg x)(z)$},
\mbox{$y' =$} \mbox{$(\neg y)(z) =$} \mbox{$(\neg y)(x)$} and
\mbox{$z' =$} \mbox{$(\neg z)(x) =$} \mbox{$(\neg z)(y)$}.
Then \mbox{$\theta(x', y', z') =$} \mbox{$- \theta(x, y, z)$}.
\end{lemma}
\begin{proof}
Choose an arbitrary unit vector $\vec{u}$ in $x$. 
Let $\vec{v}$ be the unit vector of $y$ such that 
\mbox{$\langle \vec{u} \, , \, \vec{v} \rangle > 0$}.
Let $\vec{u}'$ be the unit vector of $x'$ such that
\mbox{$\langle \vec{v} \, , \, \vec{u}' \rangle > 0$}.
Let us have \mbox{$\vec{v} =$}
\mbox{$r_{1} \vec{u} + r_{2} \vec{u}'$} for positive real numbers
\mbox{$r_{i}, i = 1 , 2$}.
The vector \mbox{$r_{2} \vec{u} - r_{1} \vec{u}'$} is a unit vector in $y'$.
Let \mbox{$\vec{v}' = r_{2} \vec{u} - r_{1} \vec{u}'$}.
Let $\vec{w}$ be the unit vector of $z$ such that 
\mbox{$\langle \vec{u} \, , \, \vec{w} \rangle > 0$}.
Let \mbox{$\vec{w} = r_{3} \vec{u} + r_{4} e^{i \varphi} \vec{u}'$}
for positive $r_{i}$'s $i = 3 , 4$ and some angle $\varphi$.
Let \mbox{$\vec{w}' = r_{4} e^{- i \varphi} \vec{u} - r_{3} \vec{u}'$},
a unit vector of $z'$.

We see that:
\[
\theta(x, y, z) \, = \, \arg(\langle \vec{u} \, , \, \vec{v} \rangle) \, + \,
\arg(\langle \vec{v} \, , \, \vec{w} \rangle) \, + \,
\arg(\langle \vec{w} \, , \, \vec{u} \rangle) \: = \:
0 + \arg(\langle \vec{v} \, , \, \vec{w} \rangle) + 0.
\]
and
\[
\theta(x', y', z') \, = \, \arg(\langle \vec{u}' \, , \, \vec{v}' \rangle) \, + \,
\arg(\langle \vec{v}' \, , \, \vec{w}' \rangle) \, + \,
\arg(\langle \vec{w}' \, , \, \vec{u}' \rangle) \, = \,
\pi + \arg(\langle \vec{v}' \, , \, \vec{w}' \rangle) + \pi.
\]
We are left to show that 
\mbox{$\arg(\langle \vec{v}' \, , \, \vec{w}' \rangle) = $}
\mbox{$- \arg(\langle \vec{v} \, , \, \vec{w} \rangle)$}.
In fact, we shall show that
\mbox{$\langle \vec{v}' \, , \, \vec{w}' \rangle = $}
\mbox{$\langle \vec{w} \, , \, \vec{v} \rangle$}.
Indeed, \mbox{$\langle \vec{v}' \, , \, \vec{w}' \rangle = $}
\mbox{$r_{2} r_{4} e^{i \varphi} + r_{1} r_{3}$}
and
\mbox{$\langle \vec{w} \, , \, \vec{v} \rangle = $} 
\mbox{$r_{1} r_{3} + r_{2} r_{4} e^{i \varphi}$}.
\end{proof}

\section{Properties of Superpositions} \label{sec:superp_prop}

A most remarkable novelty of QM is that the components of a superposition
interfere. To put this in evidence, let us consider
\mbox{$p(r y + (1 - r) z, x)$}. One would expect this quantity to be, 
essentially, equal to \mbox{$r p(y, x) + (1 - r) p(z, x)$}.
\begin{lemma} \label{le:p_basis}
If \mbox{$x \not \perp y$}, \mbox{$y \not \perp z$}, 
\mbox{$z \not \perp x$}, \mbox{$r \in [0,1]$}, 
\mbox{$\omega(r, y, z) = 1 + 2 \sqrt{r (1 -r) p(y, z)}$} we have, for any 
\mbox{$x \in X$}:
\[
p(r y \, + \, (1 - r) z, x) = 
\frac{r p(y, x) + (1 - r) p(z, x) + 
2 \cos(\theta(x, y, z)) \sqrt{r (1 - r) p(y,x) p(z, x)}} 
{\omega(r, y, z)}
\]
\end{lemma}
We see that, indeed, \mbox{$p(r y + (1 - r) z, x)$} is almost 
equal to \mbox{$r p(y, x) + (1 - r) p(z, x)$}. But there are two correction
terms. The term 
\mbox{$2 \cos(\theta(x, y, z)) \sqrt{r (1 - r) p(y,x) p(z, x)}$} is
an interference term, a characteristic of QM.
The denominator \mbox{$\omega(r, y, z)$} is a normalization factor.
Note that the interference term contains $\cos(\theta(x, y, z))$, not
$\sin(\theta(x, y, z))$. Even if all angles $\theta$ are equal to zero,
which is the case in a Euclidean space, the term is non-zero. 
\begin{proof}
Let $\vec{v}$ and $\vec{w}$ be unit vectors of $y$ and $z$
respectively with \mbox{$\langle \vec{v} \, , \, \vec{w} \rangle > 0$}.
Let \mbox{$\vec{u} = \sqrt{r} \, \vec{v} + \sqrt{1 - r} \, \vec{w}$}.
We have \mbox{$\langle \vec{v} \, , \, \vec{w} \rangle =$}
\mbox{$\langle \vec{w} \, , \, \vec{v} \rangle =$}
\mbox{$\mid \langle \vec{v} \, , \, \vec{w} \rangle \mid = $}
\mbox{$\sqrt{p(y, z)}$} and
\[
\mid \vec{u} \mid^{2} = \langle \sqrt{r} \, \vec{v} + \sqrt{1 - r} \, \vec{w} 
\, , \,
\sqrt{r} \, \vec{v} + \sqrt{1 - r} \, \vec{w} \rangle = 
\]
\[
r + 2 \sqrt{r (1 - r)} \sqrt{p(y, z)} + (1 - r) = 
\omega(r, y, z).
\]

Let now $\vec{t}$ be the unit vector of $x$ such that 
\mbox{$\langle \vec{t} \, , \, \vec{v} \rangle > 0$}. 
We have:
\mbox{$\theta(x, y, z) =$}
\mbox{$0 + 0 + \arg(\langle \vec{w} \, , \, \vec{t} \rangle)$}.
Therefore 
\mbox{$\langle \vec{w} \, , \, \vec{t} \rangle =$}
\mbox{$\sqrt{p(x, z)} e^{i \theta(x, y, z)}$} and
\[
\langle \sqrt{r} \, \vec{v} + \sqrt{1 - r} \, \vec{w} \, , \, \vec{t} \rangle =
\sqrt{r} \sqrt{p(y, x)} + \sqrt{1 - r} \sqrt{p(z, x)} (\cos(\theta(x, y, z)) 
+ i \sin(\theta(x, y, z))
\]
Therefore
\[
\mid \langle \sqrt{r} \, \vec{v} + \sqrt{1 - r} \, \vec{w} \, , \, 
\vec{t} \rangle \mid^{2}
=
(r p(x, y) + (1 -r) p(x, z) \cos^{2}(\theta(x, y, z)) + 
\]
\[
2 \sqrt{r (1 - r) p(x, y) p(x, z)} cos(\theta(x, y, z)) +
(1 - r) p(x, z) \sin^{2}(\theta(x, y, z)). 
\]
\end{proof}

\begin{lemma} \label{le:prop1}
If \mbox{$y \not \perp z$}, \mbox{$r \in [0,1]$}, and
\mbox{$x = r y \, + \, (1 - r) z$} then:
\begin{enumerate}
\item \mbox{$coplanar(x, y, z)$},
\item \mbox{$\theta(x, y, z) = 0$}, 
\item \label{form} 
\mbox{$p(x, y) = 1 - (1 - r) (1 - p(y, z)) \: / \: \left( 1 + 2 \sqrt{r (1 - r) p(y, z)} \right)$}, and
\item for any \mbox{$0 < r \leq 1$}, we have
\mbox{$p(r y + (1 - r) z, y) > p(y, z)$}.
\end{enumerate}
\end{lemma}
\begin{proof}
Let \mbox{$x = r y \, + \, (1 - r) y$} and
\mbox{$\vec{u} = \sqrt{r} \, \vec{v} + \sqrt{1 - r} \, \vec{w}$}.
Immediately, by Definition~\ref{def:superposition}, \mbox{$coplanar(x, y, z)$}.
Since \mbox{$\langle \vec{v} \, , \, \vec{w} \rangle > 0$},
we have 
\mbox{$\langle \vec{u} \, , \, \vec{v} \rangle > 0$}
and also 
\mbox{$\langle \vec{u} \, , \, \vec{w} \rangle > 0$}.
We conclude that \mbox{$\theta(x, y, z) = 0$}.
For~\ref{form}) the value of \mbox{$p(x, y)$} is straightforward from 
Lemma~\ref{le:p_basis}.
From the same Lemma,
\[
p(r y + (1 - r) z, y) = 
\frac{r + (1 - r) p(y, z) + 2 \sqrt{r (1 - r) p(y, z)}}
{1 + 2 \sqrt{r (1 - r) p(y, z)}} > 
\]
\[
\frac{r p(y, z) + (1 - r) p(y, z) + 
2 p(y, z) \sqrt{r (1 - r) p(y, z)}}
{1 + 2 \sqrt{r (1 - r) p(y, z)}} = p(y, z).
\]
\end{proof}

\begin{corollary} \label{le:co_prime}
If $x$, $x'$, $y$ and $z$ are coplanar states with \mbox{$x \perp x'$},
one has:
\[
\cos(\theta(x', y, z)) = \frac{\sqrt{p(y, z)} - \cos(\theta(x, y, z)) \,
\sqrt{p(x, y) p(x, z)}}  
{\sqrt{(1 - p(x, y))(1 - p(x, z))}}.
\] 
\end{corollary}
\begin{proof}
We have: \mbox{$p(r y + (1 - r) z, x) +  
p(r y + (1 - r) z, x') = 1$}. By Lemma~\ref{le:p_basis}:
\[
p(r y + (1 - r) z, x) = \frac{r p(x, y) + (1 - r) p(x, z) +
2 \cos(\theta(x, y, z)) \sqrt{r (1 - r) p(x, y) p(x, z)}}
{1 + 2 \sqrt{r (1 - r) p(y, z)}}
\]
and
\[
p(r y + (1 - r) z, x') = 
\]
\[
\frac{r (1 - p(x, y)) + (1 - r) (1 - p(x, z)) +
2 \cos(\theta(x', y, z)) \sqrt{r (1 - r) (1 - p(x, y)) (1 - p(x, z))}}
{1 + 2 \sqrt{r (1 - r) p(y, z)}}.
\]
Therefore
\[
1 + 2 \sqrt{r (1 - r) p(y, z)} =
\]
\[
 r + (1 - r) + 
2 \cos(\theta(x, y, z)) \sqrt{r (1 - r) p(x, y) p(x, z)} + 
\]
\[
2 \cos(\theta(x', y, z)) \sqrt{r (1 - r) (1 - p(x, y)) (1 - p(x, z))}
\]
and
\[
\sqrt{r (1 - r) p(y, z)} = 
\]
\[ 
\cos(\theta(x, y, z)) \sqrt{r (1 - r) p(x, y) p(x, z)} + 
\]
\[
\cos(\theta(x', y, z)) \sqrt{r (1 - r) (1 - p(x, y)) (1 - p(x, z))}
\]
\end{proof}

In parallel with Lemma~\ref{le:p_basis},
one would like to express \mbox{$\theta(ry + (1-r)z, x1, x2)$}
in terms of $r$ and the $p$'s and $\theta$'s of $y$, $z$, $x1$ and $x2$, for
coplanar states.
The formula obtained (by considering some orthonormal basis for the two
dimensional subspace) is, unfortunately, not very appealing and shall
not be presented here.

\section{Mappings that Preserve Superpositions} \label{sec:mappings}
It is a thesis of this paper that the structure of superpositions is 
the fundamental structure of Hilbert spaces that is meaningful for 
Quantum Physics. To support this thesis one should, now, analyze the
fundamental constructions used in Quantum Physics, such as tensor products 
and quotients as universal, i.e., categorical constructions in the
category of superposition preserving mappings.
Such an analysis has not been performed yet. Some first reflections on tensor
products may be found in Section~\ref{sec:conclusion}.

A preliminary step must be the proper definition of the category of
superposition structures and their superposition preserving mappings.
This paper does not provide for a proper definition of such a category,
whose objects must include both structures defined by Hilbert spaces,
studied here, and classical structures in which any two distinct states
are orthogonal, and all structures in-between.
We shall, therefore, consider only superposition structures defined by
some Hilbert space. A more general definition abstracting from Hilbert
spaces and based on the properties of the quantities $p$ and $\theta$
is left for future work.

Let \cH\ be a Hilbert space on the complex field, 
and $X$ be the set of all one-dimensional subspaces of \cH.
With any triple \mbox{$y, z \in X$},
\mbox{$r \in [0 , 1]$} such that \mbox{$y \not \perp z$}, we can associate
the superposition \mbox{$r y \: + \: (1 - r) z$}.
A function \mbox{$f : X_{1} \longrightarrow X_{2}$} between two such sets
of one-dimensional subspaces $X_{1}$ and $X_{2}$ preserves superpositions
iff for any \mbox{$y, z \in X_{1}$}, such that 
\mbox{$y \not \perp z$} and for any
\mbox{$r \in [0 , 1]$} the superposition, in $X_{2}$, 
\mbox{$r f(y) \, + \, (1 - r) f(z)$} is defined,
i.e., \mbox{$f(y) \not \perp f(z)$} and is equal to
\mbox{$f(r y \, + \, (1 - r) z)$}.

Note that if \mbox{$f : X_{1} \rightarrow X_{2}$} preserves superpositions 
and \mbox{$x \not \perp y$} then 
\mbox{$f(x) \not \perp f(y)$} since the superpositions
\mbox{$r f(x) \, + \, (1 - r) f(y)$} must be defined.
 
We shall now present some preliminary results concerning mappings that preserve
superpositions. First, note that if $\cH_{2}$ is a one-dimensional 
Hilbert space, then $X_{2}$ contains one element only and, for any $X_{1}$, 
the unique mapping \mbox{$X_{1} \rightarrow X_{2}$} preserves superpositions.
Such a mapping does not preserve $p$ or $\theta$.

A natural way to obtain a mapping
\mbox{$f : X_{1} \rightarrow X_{2}$} is to start from a 
{\em linear} map \mbox{$m : \cH_{1} \rightarrow \cH_{2}$}.
Such a map $m$ associates, with every one-dimensional subspace
of $\cH_{1}$, i.e., every member of $X_{1}$, 
a subspace of $\cH_{2}$ that is either one-dimensional or zero-dimensional.
Any injective, i.e., left-invertible, linear map $m$ defines an application
\mbox{$\overline{m} : X_{1} \rightarrow X_{2}$} defined by:
$\overline{m}(x)$ is the image $m(x)$ of the subspace $x$.
\begin{definition} \label{def:regular}
A mapping obtained from an injective linear mapping between 
Hilbert spaces in the way described just above will be called {\em regular}.
If such a map \mbox{$m : \cH_{1} \rightarrow \cH_{2}$} is a linear isometry, 
i.e., a unitary map of $\cH_{1}$ onto its image, we shall say that 
the mapping $\overline{m}$ is an isometry.
\end{definition}
Note that the mappings preserving superpositions described just above 
that map into a singleton are not regular unless $\cH_{1}$ is also of 
dimension one.
Note also that if \mbox{$m : \cH_{1} \rightarrow \cH_{2}$} is an
injective linear map, then, for any complex number $c$ different from zero,
the map $c \, m$ is an injective linear map 
\mbox{$\cH_{1} \rightarrow \cH_{2}$}
and that \mbox{$\overline{c \, m} = \overline{m} : X_{1} \rightarrow X_{2}$}.

We shall now characterize the regular mappings that preserve
superpositions. First, a well-known result from the theory of Hilbert spaces.
\begin{theorem} \label{the:isometry}
Let \mbox{$H_{1}, H_{2}$} be Hilbert spaces. If 
\mbox{$f : H_{1} \rightarrow H_{2}$} is a linear isometry, i.e., 
\mbox{$\parallel f(\vec{u}) \parallel =$}
\mbox{$\parallel \vec{u} \parallel$} for every \mbox{$\vec{u} \in \cH_{1}$}
then it preserves inner products:
\mbox{$\langle f(\vec{u}) \, , \, f(\vec{v}) \rangle =$}
\mbox{$\langle \vec{u} \, , \, \vec{v} \rangle$}  for every 
\mbox{$\vec{u} \, , \, \vec{v} \in \cH_{1}$}.
\end{theorem}

We now move to prove that if $\overline{m}$ is any regular mapping that
preserves superpositions, then $\overline{m}$ is an isometry.
\begin{lemma} \label{le:regular}
Let $\cH_{1}$ and $\cH_{2}$ be Hilbert spaces and let $X_{1}$ and $X_{2}$ 
be the one-dimensional subspaces of $\cH_{1}$ and $\cH_{2}$ respectively.
Assume \mbox{$m : \cH_{1} \rightarrow \cH_{2}$} is an injective linear mapping 
and that \mbox{$\overline{m} : X_{1} \rightarrow X_{2}$} 
preserves superpositions.
Then there is a strictly positive real constant $c$ such that, for every
\mbox{$\vec{u} \in \cH_{1}$}, one has \mbox{$\parallel m(\vec{u}) \parallel =$}
\mbox{$c \parallel \vec{u} \parallel$}, and $\overline{m}$ is an isometry.
\end{lemma}
\begin{proof}
Notice first that, if \mbox{$\parallel m(\vec{u}) \parallel =$}
\mbox{$c \parallel \vec{u} \parallel$} for every \mbox{$\vec{u} \in \cH_{1}$},
then, if we define \mbox{$n = m / c$} the mapping $n$ is a linear isometry and 
one has \mbox{$\overline{m} = \overline{n}$}, proving that $\overline{m}$ 
is an isometry.
 
Let $m$ be linear and assume $\overline{m}$ preserves superpositions.
Let \mbox{$x, y \in X_{1}$} be one-dimensional subspaces of $\cH_{1}$.
It is enough to show that there are unit vectors \mbox{$\vec{u}, \vec{v}$} 
in $x$ and $y$ respectively such that 
\mbox{$\parallel m(\vec{u}) \parallel =$}
\mbox{$\parallel m(\vec{v}) \parallel$}. 

If \mbox{$x = y$} the result follows from the linearity of $m$.
We may therefore assume that \mbox{$x \neq y$}.

Suppose, first, that \mbox{$x \not \perp y$}
and let \mbox{$r \in ]0, 1[$}.
There are unit vectors \mbox{$\vec{u} \in x$}, \mbox{$\vec{v} \in y$},
\mbox{$\vec{t} \in \overline{m}(x)$} and \mbox{$\vec{w} \in \overline{m}(y)$} 
such that
\mbox{$\langle \vec{u} \, , \, \vec{v} \rangle > 0$} and 
\mbox{$\langle \vec{t} \, , \, \vec{w} \rangle > 0$}.
Note that \mbox{$\overline{m}(x) \neq \overline{m}(y)$} since $m$ is injective.
The vector \mbox{$\sqrt{r} \, \vec{u} + \sqrt{1 - r} \, \vec{v}$} 
is a vector of
the superposition \mbox{$r x \, + \, (1 -r) y$}.
The vector \mbox{$\sqrt{r} \, \vec{t} + \sqrt{1 - r} \, \vec{w}$} 
is a vector of
the superposition \mbox{$r m(x) \, + \, (1 -r) m(y) =$}
\mbox{$\overline{m}(r x \, + \, (1 - r) y)$} since $\overline{m}$ preserves 
superpositions.
Since $m$ is linear the vector 
\mbox{$\sqrt{r} \, m(\vec{u}) + \sqrt{1 - r} \, m(\vec{v})$} is a vector of 
\mbox{$\overline{m}(r x \, + \, (1 - r) y)$}.
We conclude that both vectors
\mbox{$\sqrt{r} \, \vec{t} + \sqrt{1 - r} \, \vec{w}$} and
\mbox{$\sqrt{r} \, m(\vec{u}) + \sqrt{1 - r} \, m(\vec{v})$}
are members of the same one-dimensional subspace. 
This implies that \mbox{$m(\vec{u}) = d \vec{t}$} and
\mbox{$m(\vec{v}) = d \vec{w}$} for some complex number $d$ and
\mbox{$\parallel m(\vec{u}) \parallel =$} 
\mbox{$\parallel m(\vec{v}) \parallel =$}
\mbox{$\mid d \mid$}.

Let us now assume that \mbox{$x \perp y$}. We can find find some 
\mbox{$z \in X_{1}$} such that \mbox{$z \neq x$}, \mbox{$z \neq y$},
\mbox{$coplanar(z, x, y)$}. Since \mbox{$z \not \perp x$}, by the above
we can find unit vectors \mbox{$\vec{u} , \vec{w}$} in $x$ and $z$ respectively
such that \mbox{$\parallel m(\vec{u}) \parallel =$}
\mbox{$\parallel m(\vec{w}) \parallel$}. 
Similarly, we can find unit vectors 
\mbox{$\vec{v} , \vec{w'}$} in $y$ and $z$ respectively
such that \mbox{$\parallel m(\vec{v}) \parallel =$}
\mbox{$\parallel m(\vec{w'}) \parallel$}. 
But \mbox{$\parallel m(\vec{w'}) \parallel =$}
\mbox{$\parallel m(\vec{w}) \parallel$}. 
\end{proof}

We shall show now that any isometry preserves superpositions.
\begin{lemma} \label{le:unit_superp}
Let \mbox{$m : \cH_{1} \rightarrow \cH_{2}$} be a linear isometry. 
Then $\overline{m}$ preserves  $p$, $\theta$ and superpositions.
\end{lemma}
\begin{proof}
By Theorem~\ref{the:isometry}, $f$ preserves inner products and therefore 
preserves orthogonality, $p$ and $\theta$.

Assume now that \mbox{$x \not \perp y$} and
\mbox{$z = r x + (1 -r) y$}.
We have
\mbox{$\overline{m}(x) \not \perp \overline{m}(y)$} and therefore
the superposition \mbox{$r \overline{m}(x) + (1 - r) \overline{m}(y)$}
is defined.

If \mbox{$\vec{u} , \vec{v}$} are unit vectors of $x$ and $y$ respectively,
such that \mbox{$\langle \vec{u} \, , \, \vec{v} \rangle > 0$} then
\mbox{$m(\vec{u}) , m(\vec{v})$} are unit vectors of 
\mbox{$\overline{m}(x) , \overline{m}(y)$} respectively such that
\mbox{$\langle m(\vec{u}) \, , \, m(\vec{v}) \rangle > 0$} and therefore
\mbox{$r \overline{m}(x) + (1 - r) \overline{m}(y)$} is the one-dimensional
subspace generated by
\mbox{$\sqrt{r} \, m(\vec{u}) + \sqrt{1 - r} \, m(\vec{v}) =$}
\mbox{$m(\sqrt{r} \, \vec{u} + \sqrt{1 - r} \, \vec{v})$} which is
\mbox{$\overline{m}(r x \, + \, (1 - r) y)$}.
\end{proof}

We can now characterize regular mappings that preserve superpositions. 
\begin{theorem} \label{the:char_morph}
Let \mbox{$m : \cH_{1} \rightarrow \cH_{2}$} be any linear injective mapping.
The function \mbox{$\overline{m} :$}
\mbox{$X_{1} \rightarrow X_{2}$} preserves superpositions 
iff it is an isometry.
\end{theorem}
\begin{proof}
The {\em if} part is Lemma~\ref{le:unit_superp}.
The {\em only if} part is Lemma~\ref{le:regular}.
\end{proof}

\section{Conclusion and Future Work} \label{sec:conclusion}
We have shown that the properties of superpositions are governed by two
geometrical quantities $p$ and $\theta$ defined, respectively for pairs
and triples of one-dimensional subspaces in a Hilbert space, 
thus moving forward
John von Neumann's program of focusing on subspaces and not on vectors.

The most pressing task is probably now to provide an abstract definition of
structures admitting a superposition operation, 
generalizing those structures provided by Hilbert spaces.   

A quantic system composed of two sub-systems is represented by the tensor
product of the Hilbert spaces representing the two sub-systems.
Product states of the form $x_{1} \otimes x_{2}$ are elements of this tensor
product. On such product states, the quantities $p$ and $\theta$ 
are easily analyzed:
we have 
\[
p(x_{1} \otimes x_{2}, y_{1} \otimes y_{2}) =
p(x_{1}, y_{1}) p(x_{2}, y_{2})
\]
and
\[
\theta(x_{1} \otimes x_{2}, y_{1} \otimes y_{2}, z_{1} \otimes z_{2}) =
\theta(x_{1}, y_{1}, z_{1}) + \theta(x_{2}, y_{2}, z_{2}).
\]
The tensor product can be characterized as the closure of the set of product 
states under superpositions (in our sense) and the operation of taking 
the state orthogonal to a given state in a given two-dimensional plane. 

Extending this definition to superpositions of product states in accordance
with the properties of $p$ and $\theta$ on superpositions provides a 
superposition structure that is a original presentation of the tensor product
and may be found useful to study symmetry properties. 

\section{Acknowledgements} \label{sec:ack}
I am most grateful to Kurt Engesser and Dov Gabbay for extremely fruitful 
discussions during the elaboration of this paper. I thank Dorit Aharonov and
Jean-Marc L\'{e}vy-Leblond for their interest and help.

\bibliographystyle{plain}
\bibliography{../../../../my}
\end{document}